\documentstyle[11pt]{article}

\input{epsf}

\begin{document}
\renewcommand{\theequation}{\arabic{equation}}
\renewcommand{\thefigure}{\arabic{figure}}
\renewcommand{\thetable}{\arabic{table}}
\newenvironment{eqn}{\begin{equation}}{\end{equation}}
\newcommand{\beq}{\begin{eqn}}
\newcommand{\eeq}{\end{eqn}}
\vspace*{35 true mm}
\begin{raggedright}
{\Large \bf Burst Mechanisms in Hydrodynamics} \\
\large
{E. Knobloch \& J. Moehlis}\\
{\em Department of Physics, University of California, \\
Berkeley, CA 94720, USA \\
E-Mail: knobloch@physics.berkeley.edu}\\
\end{raggedright}
\section*{Abstract}
\small
\setlength{\baselineskip}{11pt}
Different mechanisms believed to be responsible for the
generation of bursts in hydrodynamical systems are reviewed and
a new mechanism capable of generating regular or irregular bursts of 
large dynamic range near threshold is described. The new mechanism is 
present in the interaction between oscillatory modes of odd and even 
parity in systems of large but finite aspect ratio, and provides an 
explanation for the bursting behavior observed in binary fluid 
convection. Additional applications of the new mechanism are proposed. \\
\setlength{\baselineskip}{12pt}

\vspace{-0.2in}
\section{Introduction}
\normalsize
Bursts of activity, be they regular or irregular, are a common occurrence in physical and biological systems. In recent years several models of
bursting behavior in hydrodynamical systems have been described using 
ideas from dynamical systems theory. In this article we review these 
and then describe a new mechanism (Moehlis \& Knobloch~\cite{moeh98a,moeh98b},
Knobloch \& Moehlis~\cite{knob99}) which provides an 
explanation for the bursting behavior observed in experiments on convection 
in $^3$He/$^4$He mixtures by Sullivan \& Ahlers~\cite{sull88}.  This 
mechanism ope-rates naturally in systems with broken D$_4$ symmetry
undergoing a Hopf bifurcation from a trivial state. This symmetry, the
symmetry group of a square, may be
present because of the geometry of the system under consideration (for 
example, the shape of the container) but also appears in large aspect 
ratio systems with reflection symmetry (Landsberg \& Knobloch~\cite{land96}). 
In either case bursting arises as a result of the nonlinear interaction 
between two nearly degenerate modes with different symmetries. 

This article is an expanded version of Knobloch \& Moehlis \cite{knob98a}.

\vspace{-0.1in}
\section{Mechanisms producing bursting}

As detailed further below, bursts come in many different forms, distinguished
by their dynamic range (i.e., the range of amplitudes during each burst),
duration, and recurrence properties. Particularly
important for the purposes of the present article is the question of whether
the observed bursts occur close to the threshold of a primary instability
or whether they are found far from threshold. In the former case a dynamical
systems approach is likely to be successful: in this regime the spatial
structure usually resembles the eigenfunctions of the linear problem and it is 
likely that only a small number of degrees of freedom participate in the 
burst. Such bursts take place fundamentally in the time domain with their 
spatial manifestation of secondary importance, in contrast to {\it pulses} 
which are structures localized in both time {\it and} space; the latter are not
considered here. Since the equations governing the evolution of primary
instabilities are often highly symmetric (see Crawford \& Knobloch 
\cite{craw91a}) global bifurcations are likely to occur and these serve as 
potential candidates for bursting mechanisms. In contrast, bursts found far 
from threshold usually involve many degrees of freedom but even here some 
progress is sometimes possible.

\subsection{Bursts in the wall region of a turbulent boundary layer
\label{section:turb_BL}}

The presence of coherent structures in a turbulent boundary layer is
well established (see, e.g., Robinson~\cite{robi91} and the collection
of articles edited by Panton~\cite{pant97}).  The space-time 
evolution of these structures is often characterized by intermittent 
bursting events involving low speed streamwise ``streaks'' of fluid.
Specifically, let $x_1, x_2$, and $x_3$ be the streamwise, wall normal, 
and spanwise directions with associated velocity components $U + u_1, u_2$, 
and $u_3$, respectively; here $U(x_2)$ is the mean flow.  In a ``burst'' the 
streak breaks up and low 
speed fluid moves upward away from the wall ($u_1<0, u_2>0$); this is
followed by a ``sweep'' in which fast fluid moves downward towards the
wall ($u_1>0, u_2<0$). After the burst/sweep cycle the streak reforms, 
often with a lateral spanwise shift.\\
\indent A low-dimensional model of the burst/sweep cycle was developed by
Aubry et al.~\cite{aubr88}; further details and later references may be 
found in Holmes et al.~\cite{holm96,holm97}.
To construct such a model the authors
used a Karhunen-Lo\`{e}ve decomposition of the data to
identify an energetically dominant empirical set of eigenfunctions, hereafter
``modes''. 
The original study of Aubry et al.~\cite{aubr88} used experimental 
data for pipe flow with $Re \sim 6750$, while later studies used 
data for channel flow from direct numerical simulation with 
$Re \sim 3000-4000$ and large eddy simulations 
with $Re \sim 13800$ (Holmes et al.~\cite{holm97}).
The model was constructed by projecting the Navier-Stokes equation
onto this basis and consists of a set of coupled ODEs for the
amplitudes of these modes. The fixed points of these equations are to
be associated with the presence of coherent structures. There are two types,
related by half-wavelength spanwise translation.  \\
\indent Numerical integration of the model 
reveals that these fixed points are typically unstable and that they are
connected by a heteroclinic cycle. In such a cycle the trajectory alternately
visits the vicinities of the two unstable fixed points.
In the model of Aubry et al. \cite{aubr88} this heteroclinic cycle
is found to be structurally stable, i.e., it persists over a {\it
range} of parameter values. This is a consequence of the O(2) symmetry
of the equations inherited from periodic boundary conditions in the
spanwise direction. Moreover, for the parameter values of
interest this cycle is {\it attracting}, i.e., it attracts all nearby
trajectories. Since the transition from one fixed point to the other 
corresponds to a spanwise translation by half a wavelength, the
recurrent excursions along such a heteroclinic cycle can be identified 
with the burst/sweep cycle described above.
However, since this cycle is attracting, the time between successive bursts 
will increase as time progresses. This is not observed and 
Aubry et al.~\cite{aubr88}
appeal to the presence of a random pressure term modeling the effect
of the outer fluid layer to kick the trajectory from the heteroclinic
cycle. In the language of Busse \cite{buss84} such a pressure
term results in a {\it statistical} $\,$ limit cycle, with the bursting events
occurring randomly in time but with a well-defined mean rate. The 
resulting temporal distribution of the burst events is characterized by
a strong exponential tail, matching experimental observations. 

Attracting structurally stable heteroclinic cycles occur in a number of 
problems of this type, i.e., mode interaction problems with O(2) symmetry 
(Armbruster et al.~\cite{armb88}, 
Proctor \& Jones~\cite{jone88}, Melbourne et al.~\cite{melb89},
Steindl \& Troger~\cite{stei96}, Krupa \cite{krup97},
Hirschberg \& Knobloch \cite{hirs98}).

\subsection{Bursts in shear flows undergoing subcritical transition to 
turbulence \label{section:subcrit}}

Experimental studies of plane Couette flow and Poiseuille (pipe) flow
have shown that at high enough values of the Reynolds number $Re$ the basic
laminar flow becomes turbulent.  However, these laminar flows are linearly 
stable at all $Re$ (see, e.g.,~Drazin \& Reid~\cite{draz81}). Consequently,
the transition to turbulence in these systems must arise from finite (i.e., 
not infinitesimal) perturbations to the basic flow; such transitions are
called subcritical since the turbulent state exists for values of $Re$ for 
which the laminar state is stable. Much recent work (reviewed in Baggett 
\& Trefethen~\cite{bagg97}) has emphasized the importance of the fact that 
the linear operators $L$ governing the evolution of perturbations of these 
flows are non-normal, i.e., $L^\dagger L \neq L L^\dagger$, where $L^\dagger$ 
is the adjoint of $L$. Linear systems with a non-normal $L$ can exhibit
transient growth even though the laminar state is linearly stable; if the
growth is large enough, nonlinearities in the system may then trigger a 
transition to turbulence.  Analysis of low-dimensional models supported by
numerical simulations suggests that the minimum perturbation amplitude 
$\epsilon$ that results in turbulence scales as 
$\epsilon ={\cal O}(Re^\alpha)$
for some $\alpha < -1$ (Baggett \& Trefethen~\cite{bagg97}). Thus, 
$\epsilon$ decreases rapidly with increasing $Re$, and the experimentally 
determined $Re$ for transition should be that value at which $\epsilon$ is 
roughly equal to perturbations due to noise or imperfections in the system. \\
\indent The turbulence excited by finite amplitude perturbations in shear flows
often takes the form of turbulent spots which can move, grow, split, and 
merge (see, e.g., Daviaud et al.~\cite{davi92}).
Turbulent spots can also burst intermittently.  For example, 
Bottin et al.~\cite{bott97} observed intermittent turbulent
bursts in plane Couette flow for $Re~\approx~325$; these bursts were triggered 
by a spanwise wire through the center of the channel, and could be localized
in space by introducing a bead into the central plane.
We focus here on this burst-like behavior, also found in low-dimensional 
models of the subcritical transition to turbulence. The important issue 
in such models is the 
attractor to which the flow settles after an appropriate finite perturbation 
to the basic laminar flow.  This will depend crucially on nonlinear terms in 
the equations, a point emphasized, for example, by 
Waleffe~\cite{wale95a,wale95b,wale97a,wale97b} and 
Dauchot \& Manneville~\cite{dauc97}.  If this attractor is a fixed point,
the system settles into a steady state in which bursts do not occur.  On the
other hand, if this attractor is a limit cycle it may be appropriate to 
interpret the resulting behavior as burst-like.  This is the case in the model
studied in Waleffe~\cite{wale95a} in which the amplitudes of streamwise
streaks, streamwise rolls, and the streak instability can periodically undergo 
short-lived explosive growths at the expense of the mean shear.  
Waleffe~\cite{wale95a,wale95b} refers to this as a self-sustaining process.
Since the laminar state is linearly stable, such limit cycles cannot bifurcate
off the laminar state.  Instead, as pointed out in 
Waleffe~\cite{wale95b,wale97a}, they may be born in a Hopf bifurcation 
(from a fixed point other than that corresponding to the laminar state) or a 
homoclinic bifurcation in which a trajectory homoclinic to a fixed point 
forms at some value of $Re$ such that for smaller (larger) values of $Re$ 
a limit cycle exists (does not exist), or vice versa.  Near such a homoclinic 
bifurcation, bursts are expected because the trajectory spends a long time 
near the fixed point, bursting away and returning in a periodic or chaotic
fashion, depending on the eigenvalues of the fixed point. Other generic 
codimension one bifurcations leading to the appearance or disappearance of 
limit cycles are saddle-node bifurcations of limit cycles and saddle-nodes
in which a pair of fixed points appears on the limit cycle (Guckenheimer 
et al.~\cite{guck97}). Limit cycles could also come in from or go off
to infinity. In other models chaotic attractors have been found
(see, e.g.,~Gebhardt \& Grossman \cite{gebh94} and Baggett 
et al.~\cite{bagg95}) and these may also give rise to burst-like behavior.

The relation of the bursts described in this section (which occur for
moderate values of $Re$, e.g., $Re \approx 325$) to those which occur in the 
turbulent boundary layer (which occur for high values of $Re$) is not 
completely clear.  However, the similarities in the phenomenology have led
Waleffe~\cite{wale97b} to propose that the 
self-sustaining process will continue to have importance in the near-wall 
region of high $Re$ flows.

\subsection{Heteroclinic connections to infinity \label{section:hci}}

Another mechanism involving heteroclinic connections, distinct from that
discussed in section \ref{section:turb_BL}, has been
investigated by Newell et al.~\cite{newe88a,newe88b} as a possible
model for spatio-temporal intermittency in turbulent flow. The authors
suggest that such systems may be viewed as nearly Hamiltonian except
during periods of localized intense dissipation. A related
``punctuated Hamiltonian'' approach to the evolution of two-dimensional 
turbulence has met with considerable success 
(Carnevale et al.~\cite{carn91} and Weiss \& McWilliams~\cite{weis93}). 
For their description Newell et al. divide the instantaneous states 
of the flow into two categories, a turbulent soup (TS) characterized by 
weak coherence, and a singular (S) state characterized by strong
coherence, and suppose that the TS and S states are generalized
saddles in an appropriate phase space. Furthermore, they suppose that in 
the Hamiltonian limit the unstable manifold of TS (S) intersects 
transversally the stable manifold of S (TS).
If the constant energy surfaces are noncompact (i.e.,~unbounded), the
evolution of the Hamiltonian system may take the system into regions
of phase space with very high (``infinite'') velocities and small
scales. These regions are identified with the S states and high
dissipation. In such a scenario the strong dissipation events are
therefore identified with excursions along heteroclinic connections to 
infinity.
Perturbations to the system (such as the addition of dissipative processes) 
may prevent the trajectory from actually reaching infinity, but this
underlying unperturbed structure implies that large excursions are 
still possible.  

Newell et al.~apply these ideas to the two-dimensional nonlinear 
Schr\"{o}dinger equation (NLSE) with perturbations in 
the form of special driving and dissipative terms which act at large and 
small scales, respectively.  Here S consists of ``filament'' solutions 
to the unperturbed NLSE which become singular in finite time and represent 
coherent structures which may occur at any position in the flow field.   
When the solution is near S a large portion of the energy is in small 
scales; for the perturbed equations the dissipative term then becomes 
important so that the filament solution is approached but collapses before 
it is reached.  This leads to a spatially and temporally random occurrence 
of localized burst-like events for the perturbed equation. 
The rate of attraction at S is determined by the faster than exponential 
rate at which the filament becomes singular, while the rate of repulsion 
at S is governed by the dissipative process and hence is unrelated to the
rate of attraction.

This bursting mechanism shares characteristics with that described in 
Kaplan et al.~\cite{kapl94} in which solutions of a 
single complex Ginzburg-Landau 
equation with periodic boundary conditions undergo faster than
exponential bursting due to a destabilizing nonlinearity and collapse
due to strong nonlinear dispersion (see also 
Bretherton \& Spiegel~\cite{bret83}).
A study of a generalization of 
Burger's equation modeling nonlocality effects suggests the presence 
of burst-like events through a similar scenario 
(Makarenko et al.~\cite{maka97}).

\subsection{Bursts in the Kolmogorov flow}

\begin{figure}[t]
\begin{center}
\epsfbox{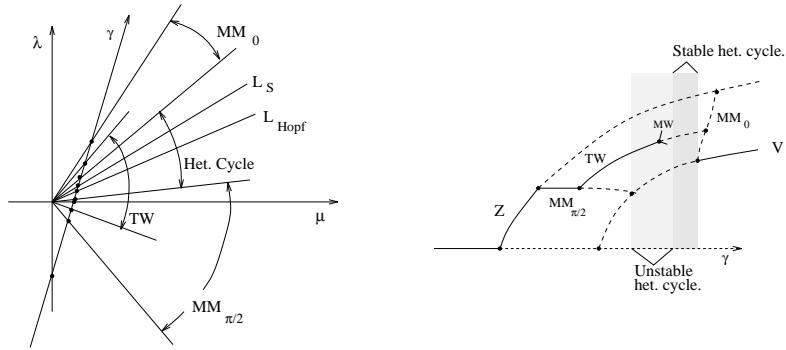}
\end{center}
\caption{Sample bifurcation diagram for the interaction of an
odd and even mode for the Kolmogorov flow obtained
by traversing the parameter space along the line $\gamma$. A
structurally stable heteroclinic cycle exists for a range
of parameters.  After Hirschberg \& Knobloch~[19].
Other cuts through parameter space give different
bifurcation diagrams.}
\label{phil} 
\end{figure}

The Kolmogorov flow ${\bf u}=(k \sin ky,0)$ is an exact solution of the 
two-dimensional incompressible Nav\-i\-er-Stokes equation 
with unidirectional forcing ${\bf f}$ at wavenumber $k$: 
${\bf f}=(\nu k^3\sin ky,0)$.  With increasing Reynolds number 
$Re\equiv\nu^{-1}$ this
flow becomes unstable, and direct numerical simulation with $2\pi$-periodic
boundary conditions shows that for moderately high Reynolds numbers
and $k>1$ the resulting flow is characterized by intermittent bursting
(She \cite{she88}, Nicolaenko \& She \cite{nico90a,nico90b,nico91,nico93},
Armbruster et al. \cite{armb92,armb94,armb96}). 
A burst occurs when the system evolves
from a coherent vortex-like modulated traveling wave (MW) to
a spatially disordered state following transfer of energy from large
to small scales. The system then relaxes to the vicinity of another
symmetry-related MW state, and the process continues with bursts occurring
irregularly but with a well-defined mean period. \\
\indent The details of what actually happens appear to depend on the value of $k$ 
because $k$ determines the symmetry of the nonlinear equation describing 
the evolution of the perturbation streamfunction $\phi$ about the Kolmogorov 
flow. Although there is no compelling reason for it, all simulations of this 
equation have been performed with $2\pi$-periodic boundary conditions in both 
directions. With these boun-dary conditions this equation has a symmetry which 
is the semi-direct product of the dihedral group D$_{2k}$ (generated by the 
actions $(x,y,\phi)\!\rightarrow (-x,-y,\phi)$ and $(x,y,\phi) \rightarrow 
(-x,y+\pi/k,-\phi)$) and the group SO(2) (representing the symmetry under 
translations $x\rightarrow x+$const). In the simplest case, $k=1$, this 
symmetry group is isomorphic to the direct product of the group O(2) of 
rotations and reflections of a circle and the group Z$_2$ representing 
reflections in the $y$-direction. Unfortunately, when $k=1$ the Kolmogorov 
flow with $2\pi$-periodic boundary conditions is not unstable for any
value of $Re$ (Meshalkin \& Sinai~\cite{mesh61}, Green~\cite{gree74}, 
Marchioro~\cite{marc86}), and one is forced to consider $k>1$.
Armbruster et al.~\cite{armb96} analyzed carefully the $k=2$ case
and showed that while a heteroclinic cycle between the MW states does form, it
is not structurally stable; the bursts are therefore not produced by
a mechanism of the type described in section \ref{section:turb_BL}. It is 
possible, however, that the onset of bursting is associated with a 
symmetry-increasing bifurcation at $Re\equiv Re_s$ (see,
e.g., Rucklidge \& Matthews~\cite{ruck96}). This would explain why the system 
stays in a single MW state for $Re$ just below $Re_s$ but visits the vicinity
of different but symmetry-related 
MW states for $Re$ just above $Re_s$.  However, despite much work a
detailed understanding of the bursts in this system remains elusive. \\
\indent An alternative approach is to consider periodic domains with different
periodicities in the two directions. In particular, if we consider the
domain  $\{-\pi<x\le\pi,-\pi/k<y\le\pi/k\}$ with $k>1$ the symmetry group 
remains O(2)$\times$Z$_2$ but
sufficiently long perturbations now grow. The unstable modes are
either even or odd under the reflection $(x,y)\rightarrow(-x,-y)$ with 
respect to a suitable origin. Mode interaction between these two steady modes 
can result in a sequence of transitions summarized in figure \ref{phil} 
(Hirschberg \& Knobloch~\cite{hirs98}): 
the Kolmogorov flow loses stability to an even mode (Z), followed
by a steady state bifurcation to a mixed parity state (MM$_{\pi/2}$). Since 
each of these states is defined to within a translation in $x$ modulo $2\pi$, 
we say that it forms a {\it circle} of states. 
The MM$_{\pi/2}$ state then loses stability in a further steady state 
bifurcation to a traveling wave (TW) which in turn loses stability at a 
Hopf bifurcation to a MW. The MW two-torus terminates in a collision with 
the two circles of pure parity states, forming an attracting structurally 
stable heteroclinic cycle connecting them and their quarter-wavelength 
translates (Hirschberg \& Knobloch~\cite{hirs98}). 
In this regime the behavior would resemble that found in the numerical
simulations, with higher modes kicking the system away from this cycle.
Indeed this sequence of transitions echoes the results obtained by 
She~\cite{she88} and Nicolaenko \& She~\cite{nico90a,nico90b,nico91,nico93}
for $k=8$.  While it is likely that the $k=1$ scenario
is relevant to these calculations because of the tendency towards an
inverse cascade in these two-dimensional systems, we emphasize that,
as mentioned above, a careful analysis of the $k=2$ case by Armbruster 
et al.~\cite{armb96}
shows that while a heteroclinic cycle of the required type does indeed form,
it is not structurally stable. The case $k=4$ has also been studied
(Platt et al.~\cite{plat91}) and a similar sequence of 
transitions found. 
Undoubtedly simulations on 
the domain $\{-\pi<x\le\pi,-\pi/k<y\le\pi/k\}$ would
shed new light on the problem, cf.~Posch \& Hoover~\cite{posc97}. \\
\indent The group O(2)$\times$Z$_2$ also arises in convection in rotating straight 
channels (Knobloch~\cite{knob96}) and in natural convection in a vertical 
slot (Xin et al. \cite{tuck98}). In both of these 
cases the linear
eigenfunctions are either even or odd with respect to a rotation by $\pi$
and the systems exhibit similar transitions. Related behavior has recently 
been observed in three-dimensional magnetoconvection with periodic boundary 
conditions on a square lattice. Unlike the Kolmogorov flow the evolution 
equation describing a steady state secondary instability of a square pattern 
is isotropic in the horizontal. Numerical simulations 
(Rucklidge et al.~\cite{ruck99}) show intermittent breakdown of a 
square pattern followed by its restoration modulo translation.

\subsection{Bursts in the Taylor-Couette system}
\label{section:TC}

\begin{figure}[h]
\begin{center}
\epsfbox{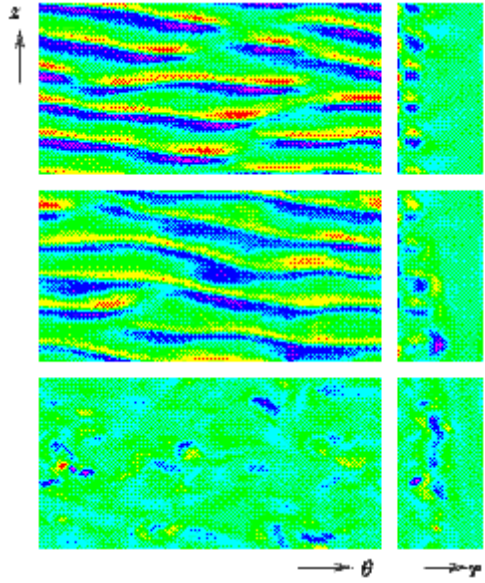}
\end{center}
\caption{
Azimuthal component of vorticity in the laminar interpenetrating spiral
state (top), just before a burst 
(middle), and in the turbulent state (bottom), labeled using polar
coordinates.
After Coughlin et al. [59].  Courtesy K. Coughlin.
}
\label{coughlin} 
\end{figure}

The Taylor-Couette system consists of concentric cylinders enclosing a
fluid-filled annulus (see, e.g., Tagg \cite{tagg94}). The cylinders can be 
rotated independently. In the
counterrotating regime the first state consists of spiral vortices of either
odd or even parity with respect to midheight. Slightly above onset the flow
resembles interpenetrating spiral (IPS) flow which may be intermittently 
interrupted by bursts of turbulence which fill the entire flow field 
(Hamill~\cite{hami95}, Coughlin et al.~\cite{coug96,coug99}). 
With periodic boundary conditions in the axial direction 
numerical simulations by Coughlin et al.~\cite{coug96,coug99} show 
that the IPS flow is temporally chaotic and consists of coexisting modes 
with different axial and azimuthal wavenumbers. This flow is confined 
primarily to the vicinity of the inner cylinder where 
the axisymmetric base flow is subject to an inviscid Rayleigh instability.
Coughlin et al. \cite{coug96,coug99} conclude that the onset of turbulence
is correlated with a secondary instability of one of the coexisting modes
of the IPS flow, namely the basic spiral vortex flow with azimuthal 
wavenumber $m=4$.  Indeed, when this mode is taken as the initial flow
for parameters chosen such that the full IPS flow undergoes bursts, a 
secondary Hopf bifurcation from this state with the same azimuthal 
wavenumber but four times the
axial wavelength is identified.  The secondary instability grows in 
amplitude and ultimately provides a finite amplitude perturbation to the 
inviscidly stable flow near the outer cylinder, triggering a turbulent 
burst throughout the whole apparatus. During a burst small scales are generated
throughout the apparatus leading to a rapid collapse of the turbulence and
restoration of the IPS flow; the bursting process then repeats roughly
periodically in time (see figure~\ref{coughlin}).  
 \\
\indent As discussed in section \ref{section:our_bursts.1}, in a finite Taylor-Couette 
apparatus there is a natural mechanism for generating bursts close to onset. 
This mechanism operates in the regime $\epsilon\sim \Gamma^{-2}$, where 
$\epsilon$ measures the fractional distance above threshold for the primary
instability and $\Gamma$ is the aspect ratio of the annulus 
(Landsberg \& Knobloch~\cite{land96}, Renardy~\cite{rena99}). 
For larger $\epsilon$ the influence of the boundaries no longer extends 
throughout the apparatus and is confined to boundary layers near the top and 
bottom. In this regime the dynamics in the bulk may be described by imposing 
periodic boundary conditions with period that is a multiple of the wavelength
of the primary instability. The success of the simulation of the observed
turbulent bursts using such periodic boundary conditions (Coughlin et al.
\cite{coug96,coug99}) suggests that these bursts occur too far above threshold
to be explained by the mechanism described in section \ref{section:our_bursts.1} 
below. This suggestion is supported by figure \ref{tc_bursts_onset} which 
compares the location of the regime where this mechanism may be expected to
operate with $\epsilon_b$, the experimental 
value of $\epsilon$ for the onset of bursts, as a function of $\Gamma$. The
latter is obtained using the approximation $\epsilon_b \approx 
(R_b - R_{IPS})/R_{IPS}$, where $R_{IPS}$ and $R_b$ are the inner cylinder 
Reynolds numbers for the onset of IPS flow and bursts, and
\begin{eqnarray*}
R_{IPS} &=& (3837 \pm 374) \Gamma^{-1} + 680 \pm 14 \\
R_b &=& (3680 \pm 374) \Gamma^{-1} + 701 \pm 14
\end{eqnarray*}
(Hamill~\cite{hami95}). This is a good approximation for strongly 
counterrotating cylinders because the IPS flow sets in for inner 
cylinder Reynolds number only about 1\% above the primary onset to spiral 
vortices (Hamill~\cite{hami95}). The figure suggests that the observed bursts
may fall within the range of validity of this theory for $\Gamma$ only
slightly smaller than those used in the experiments ($\Gamma = 17.9$,~$26$);
for such bursts the presence of endwalls should become significant.

\begin{figure}[t]
\begin{center}
\epsfbox{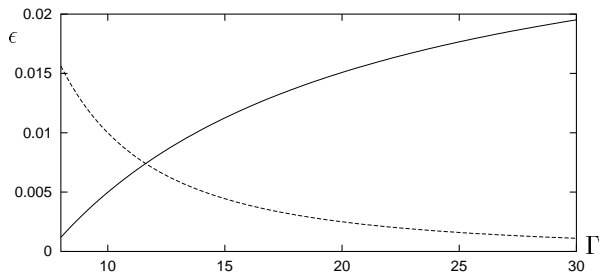}
\end{center}
\caption{
The solid line shows the inner cylinder reduced Reynolds number $\epsilon_b$ 
for the onset of bursts obtained from data in Hamill [57].  The dashed line
shows the estimate $\epsilon = \Gamma^{-2}$ near and below which the
asymptotic bursting mechanism described in section \ref{section:our_bursts.1} 
is expected to apply. The bursts found at $\Gamma = 17.9$ and $\Gamma = 26$ 
(Hamill [57], Coughlin et al. [59]) apparently fall outside the range
of validity of the theory of section \ref{section:our_bursts.1}.
}
\label{tc_bursts_onset} 
\end{figure}

\vspace{-0.1in}
\subsection{Intermittency}

The term intermittency refers to occasional, non-periodic switching between 
different types of behavior; such transitions may be viewed as bursts.
In this section we describe different types of intermittency, dividing them 
into three classes.

\subsubsection{Low-dimensional intermittency}
It is often the case that fluid dynamical systems may be modelled by
low-dimensional Poincar\'{e} maps. Fixed points of such maps correspond
to limit cycles of an appropriate dynamical system; when these lose stability
intermittency may result.  Without loss of generality we assume 
that for $\lambda<\lambda_p$ the system is in a ``laminar'' state 
corresponding to a stable fixed point of the map, and the onset of 
intermittency occurs at $\lambda =\lambda_p$.  For $\lambda>\lambda_p$ the 
state of the system will resemble the stable laminar state which existed for 
$\lambda<\lambda_p$ for some time but will intermittently undergo ``bursts'' 
away from this state, followed by ``reinjections'' to the vicinity of the 
laminar state.  The three types of intermittency identified by 
Pomeau \& Manneville \cite{pome80} correspond to three different ways in
which the laminar state ceases to exist or loses stability at
$\lambda = \lambda_p$: Type I intermittency results when a stable and 
unstable limit cycle annihilate in a saddle-node bifurcation, Type II results
when the limit cycle loses stability in a subcritical Hopf bifurcation, while
in Type III the limit cycle loses stability in an inverse period-doubling 
bifurcation. Type I and Type III intermittency have been observed, for example, 
in Rayleigh-B\'{e}nard convection (see Berg\'{e} et al.~\cite{berg80}
and Dubois et al. \cite{dubo83}, respectively), while
Type II intermittency has been observed in a hot wire experiment
(Ringuet et al.~\cite{ring93}).
A different type of intermittency, called Type X, was observed by
Price \& Mullin \cite{pric91} in a variant of the Taylor-Couette system;
this is similar to Type I intermittency but involves a hysteretic
transition due to the nature of the reinjection.  Another type of
intermittency, called Type V, was introduced in Bauer et al. \cite{baue92}
and He et al. \cite{he92}; this is similar in spirit to Type I
intermittency but involves one-dimensional maps which 
are nondifferentiable or discontinuous.  These different types of
intermittency have distinct properties, such as the scaling behavior
of the average time between bursts with $\lambda - \lambda_p$.

\subsubsection{Crisis-induced intermittency}
A crisis is a sudden change in a strange attractor as a parameter is varied
(Grebogi et al. \cite{greb82,greb83,greb87}).
There are three types of crises, and without loss of generality we assume 
that the crisis occurs as $\lambda$ is increased through $\lambda_c$.
In a boundary crisis, at $\lambda = \lambda_c$ the strange attractor 
collides with a coexisting unstable periodic orbit which lies on the
boundary of the basin of attraction of the strange attractor.  This leads
to the destruction of the strange attractor, but chaotic transients will
still exist.  Intermittency is not associated with this crisis.
In an interior crisis the strange attractor collides
with a coexisting unstable periodic orbit at $\lambda = \lambda_c$, but
here this leads to a widening rather than the destruction of the strange
attractor.  For $\lambda$ slightly larger than $\lambda_c$ the trajectory 
stays near the region of phase space occupied by the strange attractor 
before the crisis for long times, but intermittently bursts into a new region.
In an attractor merging crisis two strange attractors with basins of 
attraction separated by a basin boundary are present when $\lambda<\lambda_c$.
At $\lambda_c$ the two strange attractors simultaneously touch the basin 
boundary and ``merge'' to form a larger strange attractor for 
$\lambda>\lambda_c$.  Such a crisis is often associated with a 
symmetry-increasing bifurcation (Chossat \& Golubitsky \cite{chos88}).
For $\lambda>\lambda_c$ there is a single strange attractor on
which the trajectory intermittently switches between states resembling the
distinct strange attractors which existed for $\lambda < \lambda_c$.

\subsubsection{Intermittency involving an invariant manifold}
A manifold in phase space is invariant if every initial condition on the
manifold generates an orbit that remains on the manifold.  Invariant
manifolds often exist due to symmetries, but this is not necessary.
There are several mechanisms for intermittency involving strange
attractors on an invariant manifold.

First, suppose that as a bifurcation parameter $\lambda$ is increased through 
$\lambda_p$, the strange attractor on the invariant manifold loses stability 
transverse to the manifold; this is called a blowout bifurcation
(Ott \& Sommerer~\cite{ott94}).  Suppose that the dynamics {\it within} 
the invariant subspace do not depend on $\lambda$.  Such a system thus has 
a skew-product structure (Platt et al. \cite{plat93}), and 
$\lambda$ is called a normal parameter 
(Ashwin et al. \cite{ashw98a}).
If the blowout bifurcation is ``supercritical'' (Ott \& Sommerer~\cite{ott94})
then for $\lambda$ just above $\lambda_p$ a trajectory will spend a long time
near the invariant manifold, intermittently bursting away from it, only to 
return due to the presence of a reinjection mechanism.  This scenario is known
as on-off intermittency, where the ``off'' (``on'') state corresponds to the
system being near (away from) the invariant manifold (see, e.g.,
Platt et al. \cite{plat93}, 
Venkataramani et al. \cite{venk95}). Recently it has been shown that in
appropriate circumstances a blowout bifurcation can lead to a
structurally stable (possibly attracting) heteroclinic cycle between
chaotic invariant sets (Ashwin \& Rucklidge \cite{ashw98b}).

For systems which lack a skew-product structure and thus are governed by
non-normal parameters, in-out intermittency 
is possible (Ashwin et al. \cite{ashw98a}, Covas et al. 
\cite{cova98}).  Here the attraction and repulsion to the invariant
subspace are controlled by {\it different} dynamics.  This occurs when the
attractor for dynamics restricted to the invariant subspace is smaller
than the intersection of the attractor for the full dynamics with the
invariant subspace, and may be viewed as a generalization of on-off
intermittency in which the attraction to and repulsion from the invariant
subspace are controlled by the {\it same} dynamics.

Another mechanism for bursting occurs when the strange attractor
on the invariant manifold attracts typical orbits near the surface
but is unstable in the sense that there are unstable periodic orbits
embedded within the chaotic set which are transversely repelling.  If the
trajectory comes near such an unstable periodic orbit, there will be a
burst away from the invariant surface.  Such bursts may occur intermittently
if noise is present or small changes (called ``mismatch'') are made to the 
dynamical system that destroy the invariant surface.  The instability of 
the first of these orbits is known as a bubbling transition 
(see, e.g., Ashwin et al. \cite{ashw94}, 
Venkataramani et al. \cite{venk96a,venk96b}).

\subsection{Bursts in neural systems}

In neural systems, bursting refers to the switching of an observable such 
as a voltage or chemical concentration between an active state characterized
by rapid (spike) oscillations and a rest state.
Models of such bursting 
typically involve singularly perturbed vector fields in which system 
variables are classified as being ``fast'' or ``slow'' depending on 
whether or not they change significantly over the duration of a single 
spike.  The slow variables may then be thought of as slowly varying 
parameters for the equations describing the fast variables 
(Rinzel~\cite{rinz87a,rinz87b}, Bertram et al.~\cite{bert95}, Wang 
\& Rinzel~\cite{wang95}, Guckenheimer et al.~\cite{guck97}).
As the slow variables evolve, the state of the system in the fast variables 
may change from a stable periodic orbit (corresponding to 
the active state) to a stable fixed point (corresponding to the rest state)
and vice versa; such transitions are often associated with a region of 
bistability for the periodic orbit and the fixed point, but need not be.
Mechanisms by which such transitions can occur repeatedly have been 
classified (Rinzel~\cite{rinz87a,rinz87b}, 
Bertram et al.~\cite{bert95}, Wang \& Rinzel~\cite{wang95}).  
Behavior of the time interval between successive spikes near a transition 
from the active to the rest state is discussed by
Guckenheimer et al.~\cite{guck97}; in this paper
the presence of a subcritical Hopf-homoclinic bifurcation is also identified
as a possible mechanism for the transition from an active to a rest state.

\vspace{-0.1in}
\section{A new mechanism for bursting \label{section:our_bursts}}

\subsection{Description of the mechanism \label{section:our_bursts.1}}

We now describe a bursting mechanism which involves the interaction
between oscillatory modes in systems with approximate D$_4$ symmetry,
where D$_4$ is the symmetry group of the square.  This mechanism can lead
to bursts of large dynamic range very close to the instability
onset (Moehlis \& Knobloch~\cite{moeh98a,moeh98b}, 
Knobloch \& Moehlis~\cite{knob99}) and is expected to be relevant in many 
different systems with approximate D$_4$ symmetry. This symmetry may be 
present for obvious or subtle reasons, as the following discussion 
demonstrates. 

\begin{figure}[h]
\begin{center}
\epsfbox{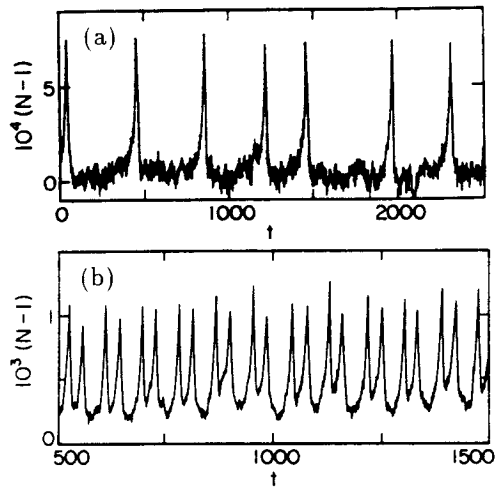}
\end{center}
\caption{
Bursts in binary fluid convection with separation ratio
$S=-0.021$. (a) $\epsilon=3 \times 10^{-4}$, 
(b) $\epsilon=3.6 \times 10^{-3}$.  The dynamic range of the bursts, measured
by the range of $(N-1)/\epsilon$ where $N$ is the Nusselt number, decreases 
with increasing $\epsilon$ while the burst frequency increases. 
After Sullivan \& Ahlers [4].  Courtesy G. Ahlers.}
\label{sullivan} 
\end{figure}

Consider binary fluid convection in a system of large but 
finite aspect ratio. If the separation ratio is sufficiently negative the 
system is overstable, i.e., the primary instability is via a Hopf bifurcation.
This is the case for the $^3$He/$^4$He mixture used by 
Sullivan \& Ahlers~\cite{sull88} in their experiment carried out in a 
rectangular container
$D \equiv \{x,y,z| -\frac{1}{2} \Gamma \le x \le \frac{1}{2} \Gamma,
-\frac{1}{2} \Gamma_y \le y \le \frac{1}{2} \Gamma_y,
-\frac{1}{2} \le z \le \frac{1}{2}\}$ with $\Gamma = 34, \Gamma_y = 6.9$.
In this experiment Sullivan \& Ahlers observed that 
immediately above threshold ($\epsilon \equiv (Ra-Ra_c)/Ra_c=3\times 10^{-4}$)
convective heat transport may take place in a sequence of irregular bursts 
of large dynamic range despite constant heat input (see figure~\ref{sullivan}).
In this system the presence of sidewalls destroys translation symmetry 
in the $x$ direction which would be present if the system were unbounded,
but with identical boundary conditions at the sidewalls the system retains
a reflection symmetry about $x=0$; the primary modes are thus either even or 
odd with respect to this reflection (Dangelmayr \& Knobloch \cite{dang91b}).
Numerical simulations of the appropriate partial differential equations
suggest that the bursts observed in the experiments
involve the interaction between the first odd and even modes of the system 
(Jacqmin \& Heminger \cite{jacq94}; see also Batiste et al.~\cite{bati98}
as described in section \ref{section:linear}).  Thus, to describe the dynamical
behavior near threshold we suppose that the perturbation from the conduction 
state takes the form
\begin{equation}
\Psi(x,y,z,t)=\epsilon^{1\over2}{\rm
Re}\;\{z_+(t) f_+(x,y,z)+z_-(t) f_-(x,y,z)\}+{\cal O}(\epsilon),
\label{perturb}
\end{equation}
where $\epsilon \ll 1$, $f_{\pm}(-x,y,z)=\pm f_{\pm}(x,y,z)$.

Following Landsberg \& Knobloch \cite{land96} we now derive amplitude 
equations describing the evolution of $z_+$ and $z_-$ using symmetry arguments.
To do this, we first briefly review the topic of bifurcations in systems
with symmetry (see, e.g., Golubitsky et al. \cite{golu88} and
Crawford \& Knobloch \cite{craw91a}).  Suppose that
\begin{equation}
\dot{v} = f(v,\lambda),
\label{equivariant}
\end{equation}
where $v \in R^n$ and $\lambda \in R^m$ represent dependent variables
and system parameters, respectively.  Let $\gamma \in G$ describe
a linear group action on the dependent variables.  We say that if
\begin{equation}
f(\gamma v,\lambda) = \gamma f(v,\lambda)
\end{equation}
for all $\gamma \in G$ then (\ref{equivariant}) is {\it equivariant} with 
respect to the group $G$.  This is equivalent to the statement that if 
$v(t)$ is a solution to (\ref{equivariant}) then so is 
$\gamma v(t)$.  For example, if a system is
equivariant under left-right reflections and a right-moving wave exists 
as a solution, then a reflection-related left-moving wave will also exist 
as a solution.
For the amplitudes $z_+$ and $z_-$ the requirement that a reflected
state (obtained by letting $x \rightarrow -x$ in (\ref{perturb})) also 
be a state of the system gives the
requirement that the amplitude equations be equivariant with respect
to the group action
\begin{equation}
\kappa_1:(z_+,z_-) \rightarrow (z_+,-z_-).
\end{equation}
Moreover, as argued by Landsberg \& Knobloch \cite{land96}, the equations for
the formally infinite system cannot distinguish between the two modes,
i.e., in this limit the amplitude equations must also be equivariant 
with respect to the group action
\begin{equation}
\kappa_2 : (z_+,z_-) \rightarrow (z_-,z_+)
\end{equation}
which we call an interchange symmetry. These two operations generate together
the symmetry group D$_4$.
For a container with large but finite length, this symmetry will be weakly
broken; in particular, the even and odd modes typically become unstable at
slightly different Rayleigh numbers and with slightly different
frequencies (see section \ref{section:linear}). The resulting equations are thus 
close to those for a 1:1 resonance, but with a special structure dictated by 
the proximity to D$_4$ symmetry. 
Finally, we may put the equations for $z_+$ and $z_-$ into normal form
by performing a series of near-identity nonlinear transformations of
the dependent variables so as to simplify the equations as much as possible
(see, e.g., Guckenheimer \& Holmes \cite{guck83}).  The normal form
equations have the additional symmetry (Elphick et al. \cite{elph87})
\begin{equation}
\hat{\sigma}:(z_+,z_-) \rightarrow {\rm e}^{i \sigma} (z_+,z_-),\qquad 
\sigma \in [0,2 \pi),
\end{equation}
which may be interpreted as a phase shift symmetry.
Truncating the resulting equations at third order we obtain
\begin{eqnarray}
\dot{z}_+ &=& [\lambda + \Delta \lambda + i (\omega + \Delta \omega)] z_+ + A (|z_+|^2 + |z_-|^2) z_+ \nonumber \\ 
&& + B |z_+|^2 z_+ + C \bar{z}_+ z_- ^2 
\label{z+}
\end{eqnarray}
\begin{eqnarray}
\dot{z}_- &=& [\lambda - \Delta \lambda + i (\omega - \Delta \omega)] z_- + A (|z_+|^2 + |z_-|^2) z_- \nonumber \\ 
&&+ B |z_-|^2 z_- + C \bar{z}_- z_+ ^2.
\label{z-}
\end{eqnarray}
Here $\Delta \omega$ measures the difference in frequency between the
two modes at onset, and $\Delta \lambda$ measures the difference in their
linear growth rates. Under appropriate nondegeneracy
conditions (which we assume here) we may neglect all interchange
symmetry-breaking contributions to the nonlinear terms. 
In the following we consider the regime in which $\lambda$,
$\Delta \lambda$, and $\Delta \omega$ are all of the same order;
in the large aspect ratio binary fluid convection context this will
occur when these quantities are all ${\cal O}(\Gamma^{-2})$ (see
section \ref{section:linear}). 
We will see that when $\Delta \lambda$ and/or $\Delta \omega$ are nonzero,
(\ref{z+},\ref{z-}) have bursting solutions.  Thus, the bursting 
mechanism may be
viewed as an interaction between spontaneous symmetry breaking (in which
the trivial conduction state loses stability to a convecting state with less
symmetry) and forced symmetry breaking (in which the presence of
sidewalls make $\Delta \lambda$ and/or $\Delta \omega$ nonzero, thereby
breaking the D$_4$ symmetry).  The introduction of small symmetry-breaking 
terms is also responsible for the possibility of complex dynamics in other 
systems that would otherwise behave in a regular manner
(Dangelmayr \& Knobloch \cite{dang87,dang91b},
Lauterbach \& Roberts \cite{laut92}, Knobloch \cite{knob96a},
Hirschberg \& Knobloch \cite{hirs96}).

To identify the bursts we introduce the change of variables
\begin{displaymath}
z_{\pm} = \rho^{-{1\over2}} \sin \left({\theta\over2}+{\pi\over4}\pm{\pi\over4}\right) \;
e^{i (\pm\phi + \psi)/2}
\end{displaymath}
and a new time-like variable $\tau$ defined by $d\tau/dt = \rho^{-1}$.
In terms of these variables (\ref{z+},\ref{z-}) become
\begin{eqnarray}
\frac{d \rho}{d \tau} &=& -\rho [2 A_R + B_R (1 + \cos^2 \theta) 
 + C_R \sin^2 \theta \cos 2 \phi] \label{rho} \\
&& - 2 (\lambda + \Delta \lambda \cos \theta) \rho^2 \nonumber \\
\frac{d \theta}{d \tau} &=& \sin \theta [\cos \theta (-B_R + C_R \cos 2 \phi) - C_I \sin 2 \phi] \nonumber \\
&& - 2 \Delta \lambda \;\rho \sin \theta \label{theta} \\
\frac{d \phi}{d \tau} &=& \cos \theta (B_I - C_I \cos 2 \phi) 
- C_R \sin 2 \phi + 2 \Delta \omega \;\rho, \label{phi}
\end{eqnarray}
where $A~=~A_R~+~iA_I$, etc.  There is also a decoupled equation for 
$\psi(t)$ so that fixed points and periodic solutions of equations 
(\ref{rho}-\ref{phi}) correspond, respectively, to periodic solutions and 
two-tori in equations (\ref{z+},\ref{z-}). 

In the following we measure the 
amplitude of the disturbance by $r\equiv |z_+|^2+|z_-|^2=\rho^{-1}$; 
thus $\rho=0$ corresponds to {\it infinite}
amplitude states. Equations (\ref{rho}-\ref{phi}) show that the restriction
to the invariant subspace $\Sigma\equiv\{\rho=0\}$ is equivalent to taking 
$\Delta\lambda=\Delta\omega=0$ in (\ref{theta},\ref{phi}). 
The resulting D$_4$-symmetric problem has three generic types of fixed
points (Swift \cite{swif88}): 

$\bullet$ $u$ solutions with $\cos \theta = 0, \cos 2 \phi = 1$ 

$\bullet$ $v$ solutions with $\cos \theta = 0, \cos 2 \phi = -1$

$\bullet$ $w$ solutions with $\sin \theta = 0$.  

\begin{figure}[t]
\begin{center}
\epsfbox{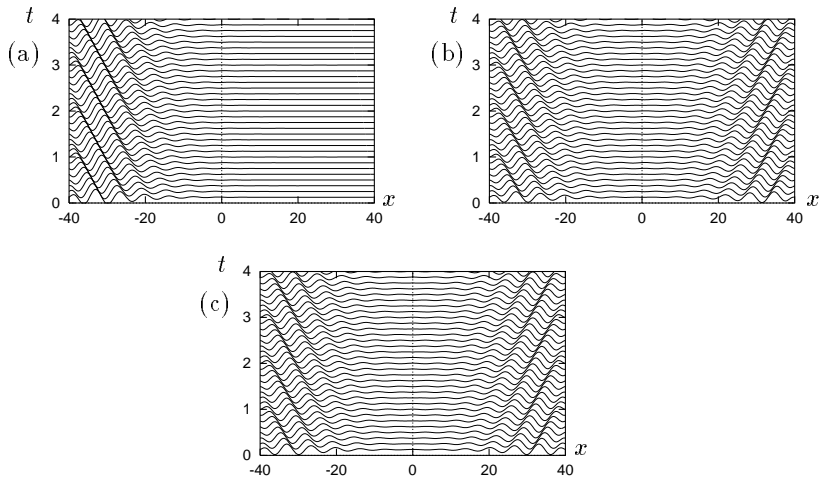}
\end{center}
\caption{
Examples of (a) $u$, (b) $v$, (c) $w$ solutions represented in a space-time 
plot showing the perturbation $\Psi$ from the trivial state.
}
\label{uvw}
\end{figure}

\begin{figure}
\begin{center}
\epsfbox{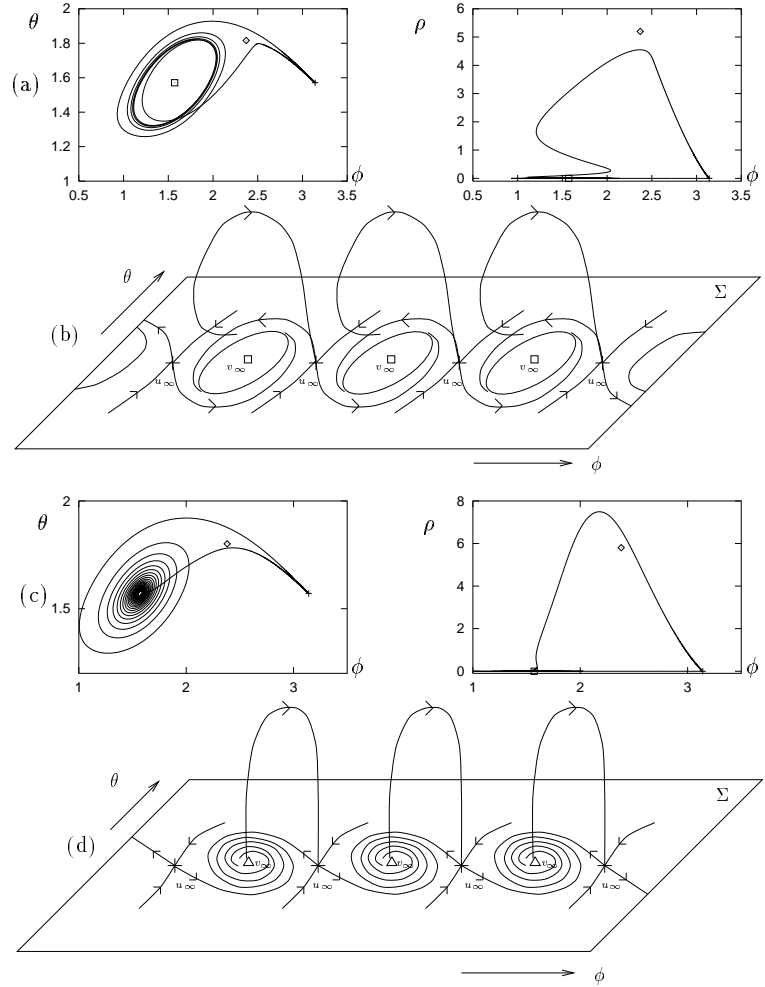}
\end{center}
\caption{
Heteroclinic cycles which exist for
$\Delta \lambda = 0.03$, $\Delta \omega = 0.02$, $A = 1 - 1.5 i$, 
$B = -2.8 + 5 i$, and (a,b) $C = 1 + i$, $\lambda=0.0974$,
(c,d) $C = 0.9 + i$, $\lambda = 0.08461$.  (a) and (c) are obtained
numerically, and (b) and (d) sketch the complete heteroclinic networks 
showing all connections.
}
\label{hetero_fig}
\end{figure}

\begin{figure}[h]
\begin{center}
\epsfbox{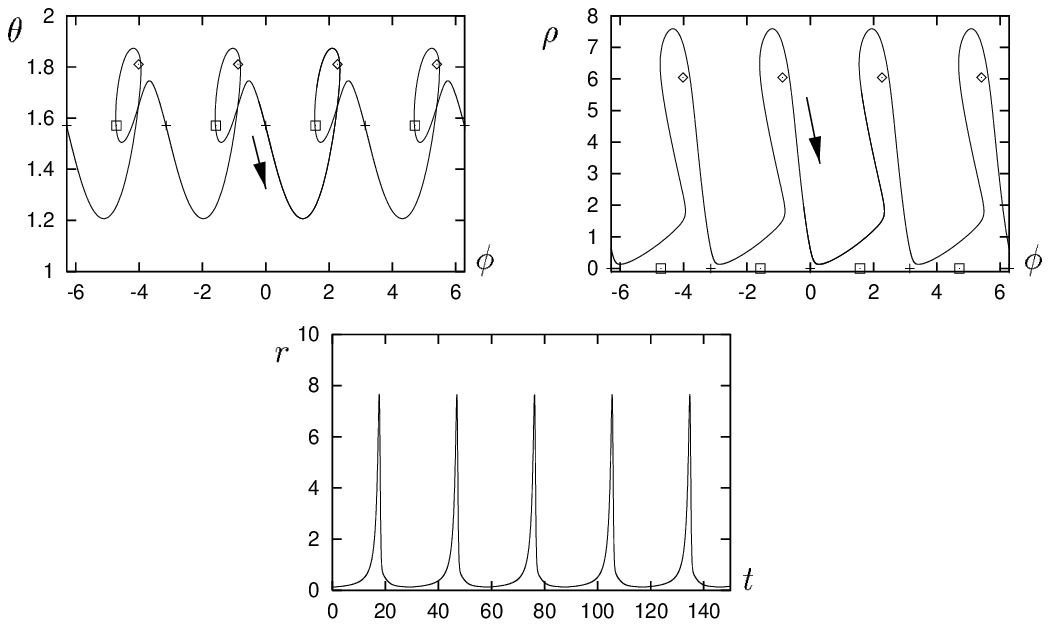}
\end{center}
\vskip -0.2in
\caption{Stable periodic rotations at $\lambda = 0.1$ for 
$\Delta \lambda = 0.03$, $\Delta \omega = 0.02$, $A = 1 - 1.5 i$, 
$B = -2.8 + 5 i$, $C =1+i$.
}
\label{rotation}
\end{figure}

\noindent
In the binary fluid context
the $u$, $v$ and $w$ solutions represent mixed parity traveling wave 
states localized near one of the container walls, mixed parity
chevron (or counterpropagating) states, and pure even ($\theta=0$)
or odd ($\theta=\pi$) parity chevron states, respectively 
(Landsberg \& Knobloch \cite{land96}).  Such states are shown in figure 
\ref{uvw} using the approximate eigenfunctions
\begin{equation}
f_{\pm}(x)=\left\{ e^{-\gamma x + i x} \pm e^{\gamma x - i x} \right\} 
\cos \frac{\pi x}{L},
\label{efunctions}
\end{equation}
where $\gamma=0.15+0.025i$, $L = 80$ and $-\frac{L}{2}\le x\le\frac{L}{2}$. 
Depending on the coefficients $A$,~$B$ and $C$ the subspace $\Sigma$ may 
contain additional 
fixed points and/or limit cycles (Swift \cite{swif88}). In our scenario, a 
burst occurs for $\lambda>0$ when a trajectory follows the stable manifold
of a fixed point (or a limit cycle) $P_1\in\Sigma$ that is {\it unstable} 
within $\Sigma$. The instability within $\Sigma$ then kicks the 
trajectory towards another fixed point (or limit cycle) $P_2\in\Sigma$. 
If this point has an unstable $\rho$ eigenvalue the trajectory escapes
from $\Sigma$ towards a finite amplitude ($\rho>0$) state, forming
a burst. If $\Delta\lambda$ and/or $\Delta\omega\ne 0$ this state
may itself be unstable to perturbations of type $P_1$ and the
process then repeats. This bursting behavior is thus associated with
a codimension one heteroclinic cycle between the {\it infinite amplitude} 
solutions $P_1$ and $P_2$ (Moehlis \& Knobloch~\cite{moeh98b},
Knobloch \& Moehlis~\cite{knob99}).  
Examples of such cycles are shown in figure \ref{hetero_fig}.  
Since in such cycles the trajectory reaches infinity in finite time the 
heteroclinic cycle actually describes bursts of {\it finite} duration
(Moehlis \& Knobloch \cite{moeh98b}). \\ 
\indent For such a heteroclinic cycle to form it is required that at least one of the 
branches in the D$_4$-symmetric system be subcritical ($P_1$) and one 
supercritical ($P_2$). Based on the $^3$He/$^4$He experiments, we focus on 
parameter values for which the $u$ solutions are subcritical and the $v$, $w$ 
solutions supercritical when $\Delta \lambda = \Delta \omega = 0$
(Moehlis \& Knobloch \cite{moeh98a}).
When $\Delta\lambda$ and/or $\Delta\omega\ne 0$, two types of oscillations 
in $(\theta,\phi)$ are possible: 

$\bullet$ rotations (see figure \ref{rotation}) 

\begin{figure}[h]
\begin{center}
\epsfbox{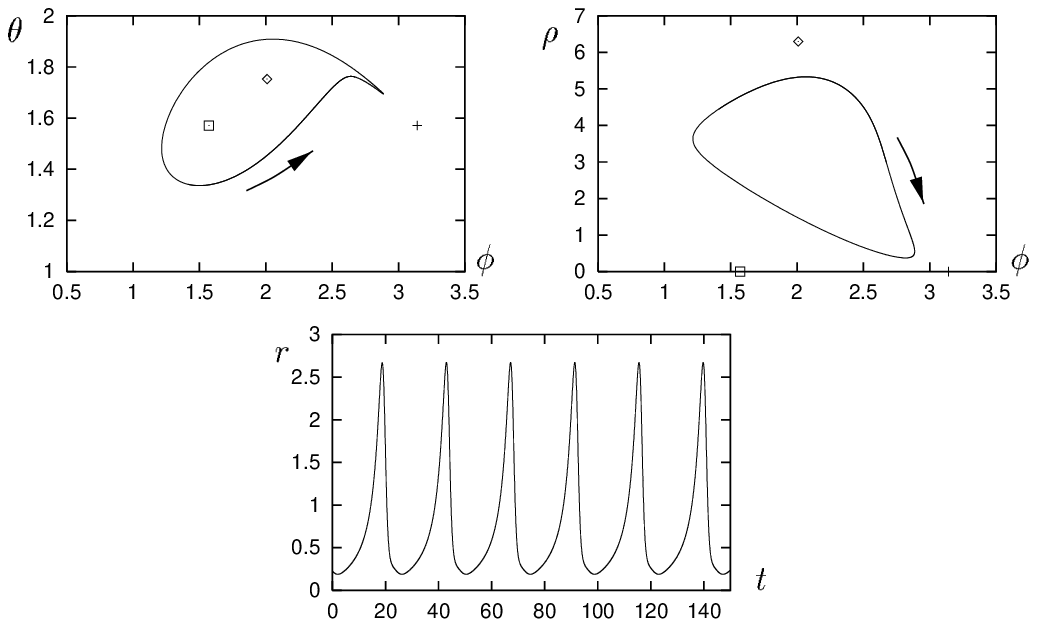}
\end{center}
\vskip -0.2in
\caption{As for figure \ref{rotation} but showing stable periodic 
librations at $\lambda = 0.1253$.}
\label{libration}
\end{figure}

$\bullet$ librations (see figure \ref{libration}). 

\begin{figure}[t]
\begin{center}
\epsfbox{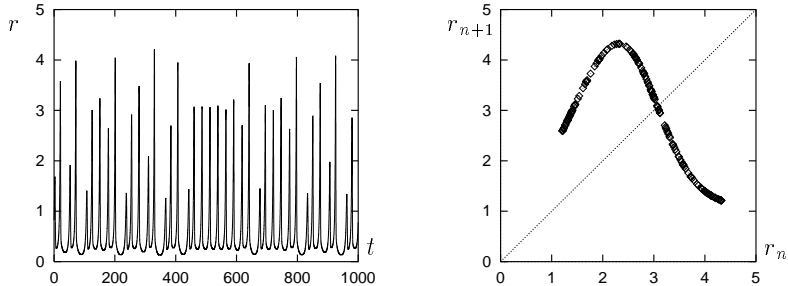}
\end{center}
\caption{Time series and peak-to-peak plot showing bursts from chaotic 
rotations with parameters as for figure \ref{rotation} but with 
$\lambda = 0.072$. 
}
\label{chaotic_rotation}
\end{figure}

\begin{figure}
\begin{center}
\epsfbox{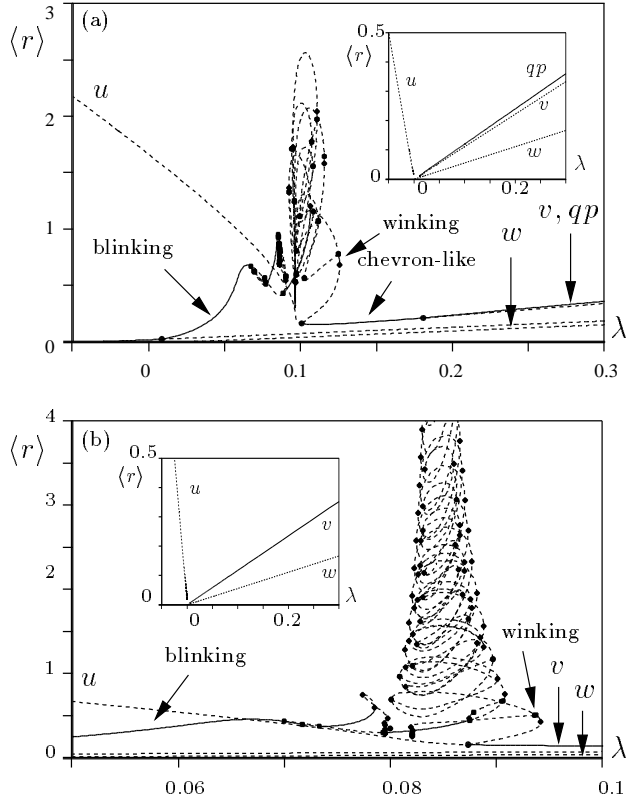}
\end{center}
\caption{Bifurcation diagrams for (a) $C= 1 + i$ and (b) 
$C= 0.9 + i$ with $A, B, \Delta \lambda, \Delta \omega$ as in figure 
\ref{rotation} showing
the time-average of $r$ for different solutions as a function of 
$\lambda$.  Solid (dashed) lines indicate stable (unstable) solutions. 
The branches labeled $u$, $v$, $w$, and $qp$ (quasiperiodic) may be 
identified in the limit of large $|\lambda|$ with branches in the 
corresponding diagrams when $\Delta\lambda=\Delta\omega=0$ (insets). All 
other branches correspond to bursting solutions which
may be blinking or winking states.  Circles, squares, and 
diamonds in the diagram indicate Hopf, period-doubling, and 
saddle-node bifurcations, respectively.}
\label{bds}
\end{figure}

\begin{figure}
\begin{center}
\epsfbox{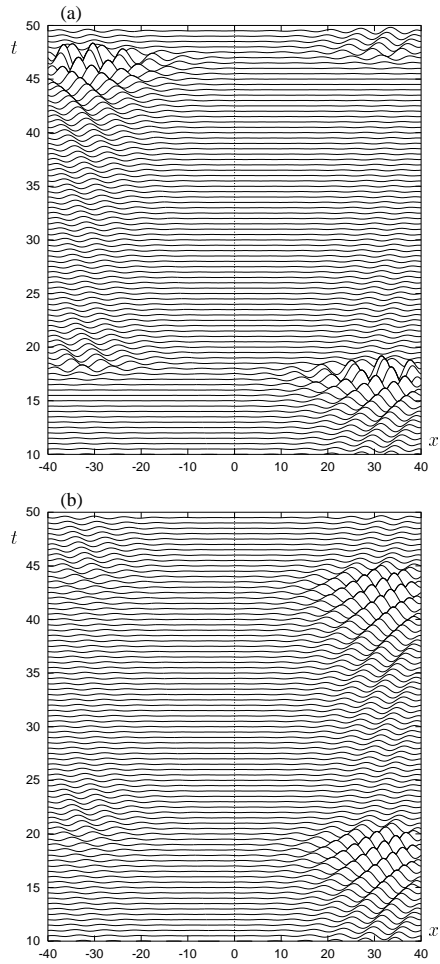}
\end{center}
\caption{The perturbation $\Psi$ from the trivial state represented in a
space-time plot showing (a) a periodic blinking state (in which successive
bursts occur at opposite sides of the container) from the trajectory
in figure \ref{rotation}, and (b) the periodic winking state (in which 
successive bursts occur at the same side of the container) for the trajectory
in figure \ref{libration}.}
\label{space_time}
\end{figure}

\begin{figure}[h]
\begin{center}
\epsfbox{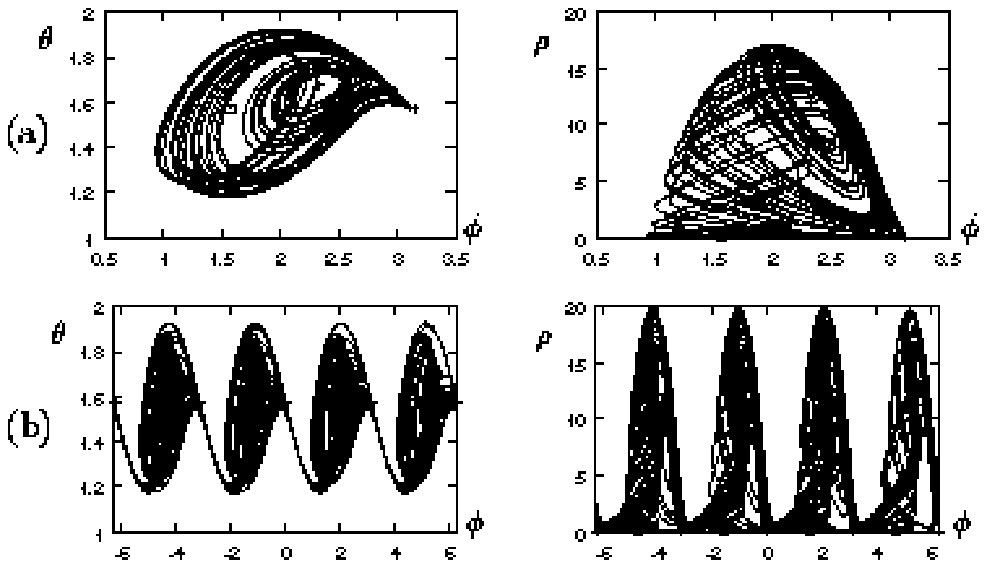}
\end{center}
\caption{
(a) Stable chaotic state with repeated visits to the 
vicinity of the same infinite amplitude state for $A = 1 - 1.5 i,\;
B = -2.8 + 5 i,\; C = 1 + i,\;\Delta\lambda = 0.03,\;\Delta \omega = -0.02$ and
$\lambda = 0.04$, corresponding to a chaotic winking state.
(b) Stable chaotic state at $\lambda=0.03$ with repeated visits to either 
the same infinite amplitude state or symmetry-related ones.  The resulting
state is neither purely blinking nor purely winking.
}
\label{new_solutions}
\end{figure}

\noindent
For $\lambda>0$ these give rise, under appropriate conditions, to sequences 
of large amplitude bursts arising from
repeated excursions towards the infinite amplitude ($\rho=0$) $u$ solutions.
Irregular bursts are also readily generated: figure \ref{chaotic_rotation} 
shows bursts arising from chaotic rotations. 
Figure \ref{bds} provides a partial summary
of the different solutions of (\ref{rho}-\ref{phi}) and their stability
properties; much of the complexity revealed in these figures is due to the
Shil'nikov-like properties of the heteroclinic cycle 
(Moehlis \& Knobloch~\cite{moeh98b}, Knobloch \& Moehlis~\cite{knob99}).
\indent We now focus on the physical manifestation of the bursts.
In figure \ref{space_time} we show the solutions of figures \ref{rotation} 
and \ref{libration} in the form of space-time 
plots using the approximate eigenfunctions (\ref{efunctions}).
The bursts in figure \ref{space_time}(a) are generated as a result of 
successive visits to {\it different} (but symmetry-related) infinite 
amplitude $u$ solutions, cf.~figure~\ref{rotation}; in 
figure~\ref{space_time}(b)
the generating trajectory makes repeated visits to the {\it same} infinite 
amplitude $u$ solution, cf.~figure~\ref{libration}. 
The former state is typical of the blinking state identified in binary 
fluid and doubly diffusive convection in rectangular containers 
(Kolodner et al.~\cite{kolo89}, 
Steinberg et al.~\cite{stei89},
Predtechensky et al.~\cite{pred94}).
It is likely that the irregular bursts reported in 
Sullivan \& Ahlers~\cite{sull88} are due to such a 
state. The latter is a new state which we call a 
{\it winking} state; winking 
\clearpage
\noindent
states may be stable but often coexist
with stable chevron-like states which are more likely to be observed 
in experiments in which the Rayleigh number is ramped upwards 
(cf.~figure \ref{bds}). 
For other values of $\Delta \lambda$ and $\Delta \omega$ it is also
possible to find stable chaotic winking states and states which are
neither purely blinking nor purely winking (see figure \ref{new_solutions}).

\begin{figure}
\begin{center}
\epsfbox{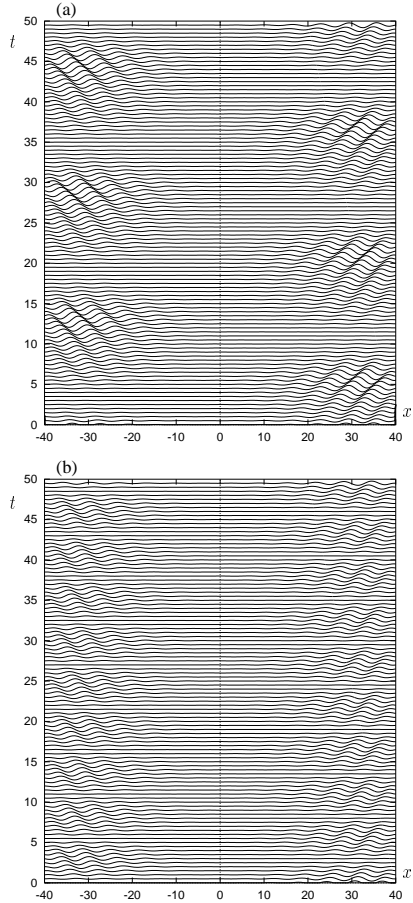}
\end{center}
\caption{The perturbation $\Psi$ from the trivial state showing stable periodic
solutions for the parameters of figure \ref{rotation} except with 
(a) $\Delta \omega = 0.1$ and (b) $\Delta \omega = 0.5$.  From these and 
figure \ref{space_time}(a) we see that the bursts fade away with increasing
$\Delta \omega$ and are replaced by smaller amplitude, 
higher frequency states.  The amplitude scales are the same here as for 
figure \ref{space_time}(a).}
\label{fade}
\end{figure}

\begin{figure}[h]
\begin{center}
\epsfbox{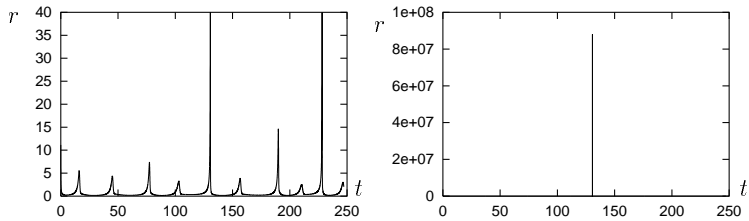}
\end{center}
\caption{A very large amplitude burst for the parameters of
figure~\ref{rotation} except with $\Delta \lambda = 0.06$,
$\Delta \omega = -0.01$.}
\label{very_large}
\end{figure}

The bursts described above are the result of oscillations in amplitude
between two modes of opposite parity and ``frozen'' spatial structure.
Consequently the above burst mechanism applies in systems in which bursts
occur very close to threshold. This occurs not only in the convection 
experiments already mentioned but also in the mathematically identical 
(counterrotating) Taylor-Couette system where counterpropagating spiral 
vortices play the same role as traveling waves in convection 
(Andereck et al. \cite{ande86}, Pierce \& Knobloch \cite{pier92}).
In slender systems, such as the convection system described above
or a long Taylor-Couette apparatus, a large aspect ratio $\Gamma$ is required 
for the presence of the approximate D$_4$ symmetry. If the size of the 
D$_4$ symmetry-breaking terms $\Delta\lambda$, $\Delta\omega$ is increased 
too much the bursts fade away and are replaced by smaller amplitude, 
higher frequency states (see figure \ref{fade}).
Indeed, if $\Delta\omega\gg\Delta\lambda$, averaging eliminates the $C$ terms 
responsible for the bursts (Landsberg \& Knobloch \cite{land96}).  
From these considerations, we conclude that bursts will not be present 
if $\Gamma$ is too small or $\epsilon$ too large. 
However, the mechanism is quite robust and even for 
$\Delta\omega\gg\Delta\lambda$ it may still be possible to choose $\lambda$ 
values so that bursts of large dynamic range occur 
(Moehlis \& Knobloch~\cite{moeh98b}). \\
\indent It is possible that the burst amplitude can become large enough 
for secondary instabilities not captured by the Ansatz (\ref{perturb}) to 
be triggered. Such instabilities could occur on very 
different scales and result in {\it turbulent} rather than just large 
amplitude bursts.  However, it should be emphasized that the physical 
amplitude 
of the bursts is ${\cal O}(\epsilon^{1\over2})$ and so approaches zero as 
$\epsilon\downarrow 0$, cf.~eq.~(\ref{perturb}).  Thus despite their large 
dynamic range (cf. figure \ref{very_large}), the bursts 
are fully and correctly described by the asymptotic expansion that leads 
to (\ref{z+},\ref{z-}).  In particular, as shown in 
Moehlis \& Knobloch~\cite{moeh98b}, the mechanism is robust with respect 
to the addition of small fifth order terms. However, the effects of including
D$_4$-symmetry-breaking terms in the cubic terms in (\ref{z+},\ref{z-}) have
not been analyzed; these terms can dominate the symmetry-breaking terms 
retained in the linear terms when $\rho$ is small.

\subsection{Applicability to binary mixtures \label{section:linear}}

In view of the motivation for studying systems with approximate D$_4$
symmetry described above, it is of interest to examine carefully the 
properties of the linear stability problem for binary fluid convection in 
finite containers. In the Boussinesq 
approximation this system is described by the nondimensionalized equations 
(Clune \& Knobloch~\cite{clun92})
\begin{eqnarray}
\partial_t {\bf u} + ({\bf u} \cdot \nabla){\bf u} & = &
-\nabla P + \sigma R [\theta(1+S) - S \eta] \hat{\bf z} + \sigma
\nabla^2
{\bf u},\\
\partial_t \theta + ({\bf u} \cdot \nabla) \theta & = & w +
\nabla^2 \theta, \\
\partial_t \eta + ({\bf u}\cdot \nabla)\eta & = & \tau \nabla^2 \eta
+ \nabla^2 \theta,
\end{eqnarray}
together with the incompressibility condition
\begin{equation} \nabla \cdot {\bf u} = 0. \label{eq:Boussinesq} \end{equation}
Here ${\bf u} \equiv (u,w)$ is the velocity field in $(x,z)$
coordinates, $P$, $\theta$ and $C$ denote the departures of the
pressure, temperature and concentration fields from their
conduction profiles, and $\eta \equiv \theta - C$. These equations are
to be solved in the rectangular domain
$D\equiv\{x,z|-\frac{1}{2}\Gamma\leq x\leq\frac{1}{2}\Gamma,
-\frac{1}{2}\leq z\leq\frac{1}{2} \}$.

The system is specified by four dimensionless parameters in addition to
the aspect ratio $\Gamma$: the separation ratio $S$, the Prandtl and Lewis 
numbers $\sigma$, $\tau$, and the Rayleigh number $R$.
The boundary conditions appropriate to the experiments are
no-slip everywhere, with the temperature fixed at the 
top and bottom and no sideways heat flux. The final set of boundary conditions
is provided by the requirement that there is no mass flux through any
of the boundaries. The boundary conditions are thus
\begin{equation}
\label{eq:bcs}
{\bf u} = {\bf n}\cdot\nabla\eta = 0 \mbox{ on $\partial D$},
\end{equation}
\begin{equation}
\label{eq:bcs1}
\theta= 0 \mbox{ at $ z = \pm 1/2$}, \qquad
\partial_x \theta = 0 \mbox{ at $x=\pm \Gamma$}.
\end{equation}
Here $\partial D$ denotes the boundary of $D$.

\begin{figure}
\begin{center}
\epsfbox{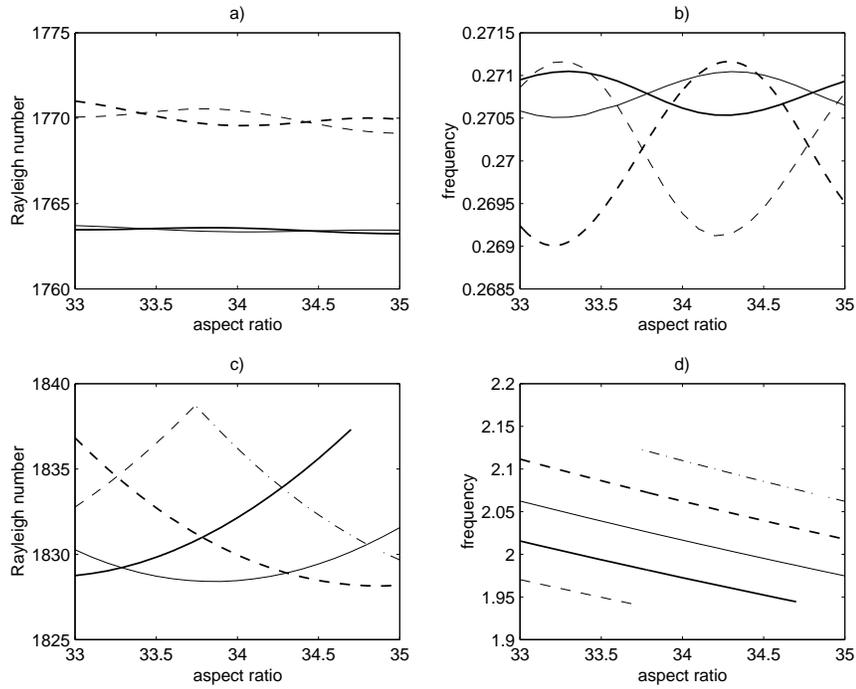}
\end{center}
\caption{Onset of convection in $^3$He/$^4$He mixtures ($\sigma=0.6$, 
$\tau=0.03$) in large aspect ratio containers. (a) Neutral stability curves
and (b) corresponding frequencies for the first two even (thick lines) and
the first two odd (thin lines) modes as a function of the aspect ratio
$\Gamma$ for $S=-0.001$. (c,d) The same but for $S=-0.021$. Courtesy O. Batiste.
}
\label{linear} 
\end{figure}

\begin{figure}
\begin{center}
\epsfbox{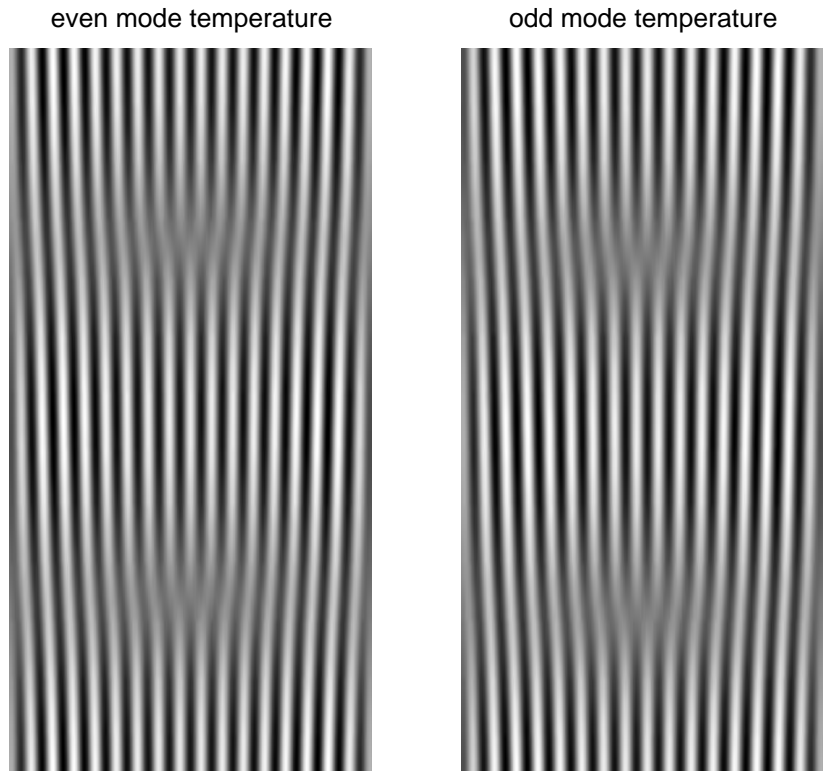}
\end{center}
\caption{The first even and odd temperature eigenfunctions when
$\sigma=0.6$, $\tau=0.03$, $\Gamma=34.0$ and $S=-0.021$ in the form of a 
space-time plot with time increasing upward. Courtesy O. Batiste.
}
\label{efunction_plot} 
\end{figure}

Figure \ref{linear} shows the results of solving the {\it linear} problem 
describing the stability properties of the conduction state 
${\bf u}=\theta=\eta=0$ for parameter values used by 
Sullivan \& Ahlers~\cite{sull88} in their $^3$He/$^4$He 
experiment: $\sigma=0.6$, $\tau=0.03$, $\Gamma=34.0$.
The figure shows the neutral stability curves and corresponding frequencies 
for the first {\it four} modes in the range $33.0\le\Gamma\le 35.0$ 
for $S=-0.001$ and $S=-0.021$. Observe that when $|S|$ is sufficiently 
small the first two families of neutral curves are separated by a gap that is 
much larger than the amplitude of the ``braids'' within each family 
(figure \ref{linear}(a)).  This is typical of what happens in 
Rayleigh-B\'{e}nard convection with non-Neumann boundary conditions 
(Hirschberg \& Knobloch \cite{hirs97}) and makes it
easy to justify projecting the fluid equations onto the first two modes
that become unstable. However, the situation is not so simple. This is
because in the case of overstability this behavior does not persist for all 
$\Gamma$ or all values of $|S|$. For larger values of these parameters the 
results take instead the form shown in figures \ref{linear}(c,d) which
show the linear stability results for $S=-0.021$ and the same range 
of values of $\Gamma$ as figures~\ref{linear}(a,b).
The modes from the different families now cross and the first unstable mode
belongs to successively higher and higher families when extrapolated to small
$|S|$ (Batiste et al. \cite{bati98}). Figure~\ref{linear}(c) shows 
the crossing of two even modes involving a {\it nonresonant} double Hopf 
bifurcation (figure~\ref{linear}(d)). As discussed in detail in 
Batiste et al. \cite{bati98} the transition between these two types 
of behavior is mediated by a {\it resonant} 1:1 mode crossing at a somewhat 
smaller value of $|S|$. The experimental value of the separation ratio
from Sullivan \& Ahlers~\cite{sull88}, $S=-0.021$, therefore corresponds 
to the ``crossing'' case and the projection of the equations onto two modes 
cannot be rigorously justified except in the neighborhood of mode crossing 
points. \\
\indent We denote the growth rates and frequencies of the modes $z_{\pm}$ by
$\lambda_\pm$ and $\omega_\pm$. For large aspect ratios, the mode frequencies
must go like $\omega_{\pm}\sim\omega_{\infty}+c_{1\pm}\Gamma^{-1}+
c_{2\pm}\Gamma^{-2}+\cdots$. The fact that the frequency curves in 
figure~\ref{linear}(d) are essentially parallel ``straight lines'' implies that
$c_{1+}\approx c_{1-}$.  Therefore, $\Delta\omega\equiv (\omega_+-\omega_-)/2 =
{\cal O}(\Gamma^{-2})$ for large $\Gamma$ (Batiste et al.~\cite{bati98}).
Moreover, as argued in Landsberg \& Knobloch~\cite{land96}, the parabolic 
minimum of the neutral stability curve leads to the expectation that 
$\Delta \lambda = {\cal O}(\Gamma^{-2})$.
Thus, in the range $\lambda = {\cal O}(\Gamma^{-2})$, $\lambda$, 
$\Delta \lambda$, and $\Delta\omega$ are all of the {\it same} order as 
$\Gamma\rightarrow\infty$, as required for the applicability of equations 
(\ref{z+}, \ref{z-}). Of course, close enough to the mode 
crossing point $\Delta \lambda\ll\Delta\omega$, and in this region averaging 
methods can be used to eliminate the $(\bar{z}_+z_-^2,\bar{z}_-z_+^2)$ terms 
from the mode interaction equations (Landsberg \& Knobloch \cite{land96}). 
However, for typical values of $\Delta \lambda$ it appears likely that the 
system is correctly described by equations (\ref{z+}, \ref{z-}), as 
hypothesized in Landsberg \& Knobloch \cite{land96} and Moehlis \& Knobloch
\cite{moeh98a}. \\
\indent The first odd and even temperature eigenfunctions for $S=-0.021$ are shown in 
figure~\ref{efunction_plot} in the form of a space-time diagram, with time 
increasing upward. 
As in the approximate expression (\ref{efunctions}) the 
eigenfunction consists of waves propagating outwards from the center of the 
container.  The eigenfunction amplitude has a local minimum at the center 
and increases outwards, peaking near the sidewalls. This type 
of eigenfunction was anticipated by Cross \cite{cros86} and is characteristic 
of eigenfunctions in systems with {\it positive} group velocity (although,
strictly speaking, in a finite system one cannot define a group velocity 
since the allowed wave number is quantized by the sidewalls as well as being 
nonuniform). However, for the present purposes the most important observation 
is that for aspect ratios as large as this, the odd and even eigenfunctions are
essentially indistinguishable, as hypothesized by 
Landsberg \& Knobloch~\cite{land96}. 

\subsection{Other systems with approximate D$_4$ symmetry 
\label{section:other_D4}}

There are a number of other systems of interest where an approximate D$_4$ 
symmetry arises in a natural way and the bursting mechanism 
described in section \ref{section:our_bursts.1} may be relevant. These
include overstable convection in small aspect ratio containers with nearly 
square cross-section (Armbruster \cite{armb90,armb91a}) 
and more generally any partial differential equation on a nearly square domain 
describing the evolution of an oscillatory instability, cf.
Ashwin \& Mei~\cite{ashw95}.
Other systems in which our bursting mechanism might be detected are
electrohydrodynamic convection in liquid crystals (Silber et al.~\cite{silb92};
T. Peacock, private communication),
lasers (Feng et al.~\cite{feng94}), spring-supported 
fluid-conveying tubes (Steindl \& Troger~\cite{stei95}),
and dynamo theories of magnetic field generation in the Sun 
(Knobloch \& Landsberg~\cite{knob96b}, 
Knobloch et al.~\cite{knob98b}). 

Perhaps more interesting is the possibility that large scale
spatial modulation due to distant walls may produce bursting in a fully
nonlinear state with D$_4$ symmetry undergoing a symmetry-breaking Hopf 
bifurcation. As an example we envisage a steady pattern of fully nonlinear 
two-dimensional rolls. With periodic boundary conditions with period four
times the basic roll period, the roll pattern has D$_4$ symmetry since the
pattern is preserved under spatial translations by 1/4 period and a 
reflection. If
such a pattern undergoes a secondary Hopf bifurcation with a {\it spatial} 
Floquet multiplier $\exp i\pi/2$, the Hopf bifurcation breaks D$_4$
symmetry. If the invariance of the basic pattern under translations by
1/4 period is only approximate (this would be the case if the roll amplitude
varied on a slow spatial scale), the D$_4$ symmetry itself would be weakly
broken and the new mechanism described above could operate.

Also of interest is the Faraday system in a nearly square container.  
In this system gravity-capillary waves are excited on the surface of a 
viscous fluid by vertical vibration of the container, usually as a result of a
subharmonic resonance. Simonelli \& Gollub \cite{simo89} studied the
effect of changing the shape of the container from a square to a slightly
rectangular container, focusing on the $(3,2)$, $(2,3)$ interaction in this
system. These modes are degenerate in a square container and only pure and
mixed modes were found in this case. In a slightly rectangular container the
degeneracy between these modes is broken, however, and in this case 
a region of quasiperiodic and chaotic behavior was present near onset.
When these oscillations first appear they take the form of relaxation
oscillations in which the surface of the fluid remains flat for a long
time before a ``large wave grows, reaches a maximum, and decays, all in a time
short compared with the period''. The duration of the spikes is practically
independent of the forcing amplitude, while the interspike period appears
to diverge as the forcing amplitude {\it decreases}. The spikes themselves
possess the characteristic asymmetry seen in figures~\ref{rotation} and
\ref{libration}. This behavior 
occurs when the forcing frequency lies {\it below} the resonance frequency of
the square container, i.e., precisely when D$_4$-symmetric problem has 
a {\it subcritical} branch. Irregular bursts are also found, depending on 
parameters, but these are distinct from the chaotic states found by 
Nagata \cite{naga91} far from threshold and present even in a square container.

\section{Discussion}

In this article we have seen that there are many different mechanisms 
responsible for bursting in hydrodynamical systems. The table below 
summarizes the different mechanisms described in terms of properties that 
are most relevant to hydrodynamics. Thus no single mechanism can be 
expected to provide a universal explanation for all observations.
The bursts found experimentally in Taylor-Couette flow (cf. section 2.5)
and large aspect ratio binary fluid convection (cf. section 3.1)
occur very close to the threshold of a primary instability and thus have the
greatest potential for a successful dynamical systems interpretation of the
type emphasized here.  We have seen, however, that even for fully developed 
turbulent boundary layers at very large Reynolds numbers a dynamical systems
approach can be profitable (cf. section 2.1). \\
\indent Although nearly all of the mechanisms we have described rely on the presence 
of global bifurcations, there are important differences among them.  For 
example, the bursts in the wall region of a turbulent boundary layer described
in section \ref{section:turb_BL} are due to a (structurally stable) heteroclinic 
cycle connecting fixed points with {\it finite} amplitude; such a cycle 
leads to bursts with a limited dynamic range. In contrast, in the 
mechanism of section \ref{section:our_bursts.1} the dynamic range is 
unlimited. Moreover, the role of the fixed points is different: in the former
the bursts are associated with the {\it excursions} between the fixed points
while in the latter the bursts are associated {\it with} the fixed points.
Because of the asymptotic stability of the cycle the time between successive 
bursts in the turbulent boundary layer will increase without bound unless the
stochastic pressure term is included; such a stochastic term is not required
in the mechanism of section \ref{section:our_bursts.1}. In particular, in this 
mechanism the duration
of the bursts remains finite despite the fact that they are associated
with a heteroclinic connection. This is because of the faster than 
exponential escape to ``infinity'' that is typical of this mechanism. This is
so also for the mechanism described in section \ref{section:hci}. Both of these
mechanisms involve global connections to infinity and hence are capable of 
describing bursts of arbitrarily large dynamic range. The 
models of the subcritical transition to turbulence and various types 
of intermittency also produce bursts of finite duration but rely on
global reinjection which produces bursts of bounded amplitude.

\vskip 0.1in

This work was supported by NSF under grant DMS-9703684 and by NASA under 
grant NAG3-2152.

\begin{small}
\begin{table}[b]
\begin{center}
\begin{tabular}{|c|c|c|c|c|}
\hline
 & & burst & dynamic \\
section & mechanism & recurrence & range \\
 & & properties & \\
\hline\hline
2.1 & structurally stable, & increasing period &\\
       & attracting & in absence & finite\\
       & heteroclinic cycle & of random &\\
       & & pressure force & \\
\hline
2.2 & periodic orbit & periodic &\\
       & near homoclinic & or chaotic & finite\\
       & bifurcation & &\\
\hline
2.3 & heteroclinic & &\\
       & connection & ``random'' & unlimited\\
       & to infinity & &\\
\hline
2.4 & & irregular but &\\
       & symmetry-increasing & with well- &finite\\
       & bifurcation? & defined mean &\\
       & & period &\\
\hline
2.5 & finite amplitude & &\\
       & trigger to inviscidly & roughly & finite \\
       & stable flow from & periodic &\\
       & secondary instability & &\\
\hline
2.6.1 & laminar state ceases & &\\
         & to exist or loses & chaotic & finite \\
         & stability & &\\
\hline
2.6.2 & crisis of strange & chaotic & finite\\
         & attractor & &\\
\hline
2.6.3 & loss of stability out & chaotic & finite\\
         & of invariant manifold & &\\
\hline
2.7 & slow variables cause & &\\
         & effective change in & periodic & finite\\
         & parameters causing & or chaotic&\\
         & change in state & &\\
\hline
3.1 & heteroclinic cycles & periodic &\\
       & involving infinite & or chaotic & unlimited\\
       & amplitude states & &\\
\hline

\end{tabular}
\vskip0.1in\noindent
\end{center}
Table 1. The different mechanisms discussed in this article. The term
dynamic range refers to the range of amplitudes during each burst.
\end{table}
\end{small}

\end{document}